# Studies on European beech (*Fagus sylvatica* L.). Part 1: Variations of wood colour parameters

Shengquan LIU, Caroline LOUP, Joseph GRIL, Olivier DUMONCEAUD, Bernard THIBAUT

* Laboratoire de Mécanique et Génie Civil, Université Montpellier 2, France

tel: 04 6714 3433

fax: 04 6714 4792

jgril@lmgc.univ-montp2.fr

**Short Title**

Colour variations of European beech wood

## Abstract

Colour parameters of European beech were measured using CIELab system. 103 logs from 87 trees in 9 sites were cut into boards to study the variations of wood colour parameters. Both site and tree effect on colour were observed. Patterns of red heartwood occurrence were defined. When excepting red heartwood there was still a highly significant effect of site and tree; differences remained after veneer processing. Axial variations were small, except very near the pith or in red heartwood, suggesting possible early selection at periphery under colour criteria. Red heartwood is darker, redder and more yellow than normal peripheral wood.

## Résumé

Les paramètres de couleur du Hêtre Européen ont été mesurés à l'aide du système CIELab. Cent trois grumes obtenues à partir de 87 arbres abattus dans 9 sites ont été débitées en quartiers afin de déterminer les variations radiales des paramètres de couleur. Des effets site et arbre sur la couleur ont été observés. Des types avec ou sans cœur rouge ont été définis. En exceptant le cœur rouge on conserve un effet hautement significatif du site et de l'arbre. Les variations radiales et axiales étaient faibles, excepté très près de la moelle ou dans le cœur rouge, suggérant la possibilité d'une sélection précoce sur des critères de couleur en périphérie. Le cœur rouge est plus foncé, plus rouge et plus jaune que le bois normal périphérique.

**Mots-clés**

*Fagus sylvatica* **L. / système de couleur CIELab / bois matériau / cœur rouge**

# Introduction

The colour of wood differs widely among species and also within a tree. It is an important factor for end user to consider and the price of wood is often dependent on its colour parameters [2, 3]. European beech (*Fagus sylvatica* L.) is a popular and major tree species distributed in the whole Europe. Its timber with beautiful grain and proper texture is widely used in sawing, veneer, decoration and furniture. In western Europe beech is appreciated for its light pinkish colour: darker wood is less valuable in general. Moreover, industrial operations using heat treatment such as steaming or hot drying are known to change beech colour by inducing a more or less pronounced reddening and darkening. Although process parameters are of the outmost importance in these phenomena, it should be interesting to know how beech wood colour is dependent on intra or inter trees, intra or between sites variations. In the present paper, the variations of colour parameters from pith to bark were studied in different trees from 9 sites under different growth conditions and management practices.

# Material and Methods

## Material

87 trees of European beech were taken from 9 European sites from 5 different countries (Austria, Denmark, France, Germany and Switzerland, designated by A, D, F, G and S respectively), with 9-10 trees selected per site. One log of 50cm long at the height of 4m (bottom) was cut for each tree. In addition, another log was cut at the height of 9m (top) in only 9 selected trees to compare the wood properties between top and bottom of the stem. The age of the selected trees ranges from 70 to 200, the diameter at breast height (DBH) from 51 to 85cm and the tree height from 30m to 43m (table 1). Data for top logs are given in table 2. The "red heartwood" proportion defined as the ratio of red heart zone diameter to log diameter, is also indicated for each log.

Stands A1, D1, F1, G1 and S1 (left column in the table) belonged to a first campaign where more data were measured. These 5 stands were chosen for their similar growing condition: rather low altitude (about 500m above see level), typical high forest with rather narrow spacing. Based on the results from the first campaign, the procedure was somewhat simplified for stands A2, S2, G2 and G3 from the second campaign (right column), as will be explained later. These stands were selected to represent specific situations encountered in the various countries: mountain forest for A2 and S2, with a pronounced slope, high forest with large spacing for G2 and G3. The stand G3 contained two age classes (220-230 years and 140-160 years) and allowed very large spacing between trees, thus holding some similarities with typical French middle forests ("taillis sous futaie"). The harvesting occurred between November 1998 to February 1999 for the first campaign, between October 1999 and January 2000 for the second.

Each log was sawn into two radial boards through pith from north to south direction, labelled N and S, respectively. These boards were dried under shelter in open air during several weeks until they reach a moisture contant of 12 to 14 %, than planed. Colour measurement was performed immediatly after planing in order to ovoid any aging of the surface.

The colour parameters of boards were measured every 1 cm from pith to bark (figure 1).

**Colour measurement**

The measurements of colour parameters were performed in the wood physics laboratory of CIRAD with a spectrocolorimeter (Datacolor Microflash 200d) under ambient temperature and humidity from July to September 1999 (1st campaign) and in october 2000 (2nd campaign) The diameter of sensor head was 6 mm (SAV, "small area view"), the illuminant A and 10° standard observer were used as the conditions of measurement [3]. We obtained the values of the CIELAB colour system (L*, a* and b*) directly, in which L* means brightness, a* means red colour, b* means yellow colour. A larger L*, a* or b* means a lighter, redder and more yellow colour, respectively [4, 8]. Occasionally, the following colour parameters derived from L*,a*,b* will be documented:

$$C^* = (a^{*2}+b^{*2})^{1/2}$$

$$H^* = \mathrm{atan}(b^*/a^*)$$

In each pair of board the width of the red heart zone was measured; the ratio between this width and total diameter was used as an indicator of red heartwood occurrence in the log.

## Results and Discussion

**General results**

In general, 17 to 33 points were measured along a radius according to the tree diameter. The lightness index (L*) ranged from 58.2 to 90.3, redness index (a*) from 6.2 to 18.7, yellowness index (b*) from 15.4 to 30.3, C* from 16.9 to 33.4 and H* from 50.4° to 72.5°. In figure 2 beech colour is compared to that of various hardwood species [7]: it is characterised by a high lightness L*. The mean and standard deviation obtained in the present set of data has been indicated by segments, as well as results of red hardwood beech that will be discussed later. Mean values obtained on Oriental beech (*Fagus orientalis*) [6] are also shown for comparison.

There is a highly significant correlation between L* and a* (L*= -2.37a*+103.8 with $R^2$ = 0.82) and a lower correlation between L* and b* (L*=-1.88b*+118.0 with $R^2$= 0.49) or a* and b* (b*=0.80a*+12.4 with $R^2$ = 0.67, number of couples = 4548). The lighter the wood, the less red and less yellow it is. However, the scatter is considerable especially in the relation between L* and b*.

A major defect in Beech wood products is the presence of red heartwood, so it had to be characterised. This was done by observing, for each board, the variation patterns of colour parameters from pith to bark. Figure 3 shows typical patterns observed on pairs of opposite boards. Usually, there is a similar variation pattern of colour parameters from pith to bark in the north and in the south directions for the same log: either no or very few radial variations like in (a) or a pronounced change in the central portion of the stem like in (b), indicating the presence of red heart. The irregular case illustrated by (c) can be partly attributed to the eccentricity of the stem that prevented the symmetric cutting of the two boards. The two boards of (a) and the southern board of (c) are examples of "non-red heart" (NRH); whereas the two boards of (b) and the northern board of (c) will be classified as "red heart" (RH) boards.

**Accounting for the Red Heartwood**

To compare the colour of NRH and RH, we selected 5 successive positions near the pith, usually points 5 to 9 except in the case of red heart where we adjusted to the position of the darkest zone. The examples of selected zones are indicated by rectangles in figure 3. It was also necessary to compare these NRH or RH to the wood close to the periphery. For each board, we selected the last 5 points (before the very last) nearest to the bark, as outer wood. These positions will be labelled as "peripheral wood" (PW). In tables 1 and 2 the logs containing red heart were indicated by a non-zero value of redheart diameter $D_{RH}$. The lowest values correspond to the case of boards containing too few red heart positions to be classified as RH boards. In addition a "type of board" column indicates with "R-R" a log with two RHW boards, with "N-N" a log with two NRH boards and with "R-N" a log where both boards are of different type.

Figure 4 shows the relation between the colour of PW and corresponding RH or NRH, depending on the case; each point is obtained by averaging the 5 selected positions. RH is systematically darker (9.6 +/- 3.6), redder (3.9 +/- 1.3) and more yellow (2.7 +/- 1.7) than PW, while no obvious difference is observed between NRH and PW, except for a few boards exhibiting very low level of PW redness. The relation between colour parameters is shown in figure 5, separating the means of PW, RH and NRH for all boards. The considerable scatter in the relationship between L* and b* and between a* and b*, are tendencies that would have been observed when considering the whole range of values. RH forms a clearly distinct group, while NRH and PW are difficult to distinguish, except for NRH being slightly redder than PW. Red heartwood appears more or less in the continuity of normal wood in each case. Variance analysis indicated that the colour difference between RH and PW is significant at the 0.1% level for all colour parameters. Between NRH and PW the difference of a* is significant at the 0.1 level, that of L* at the 5% level; that of b* is not significant even at the 5% level.

At this point, a comment can be made concerning the denomination of red heartwood. In the case of a board with red heart, this higher level of redness in the heart, a tendency already observed in normal situation, is exacerbated. However, the systematic observation of radial profiles suggested that the transition between the so-called "red heartwood" zone and the "normal" zone is often sharper with respect to lightness than to redness. Therefore, the qualification of "dark" can be as appropriate as that of "red".

**Variations among trees and stands**

There exist significant differences among the 9 stands and among trees in each stand. The grouping of trees or stand is not always the same depending on the chosen colour parameter.

Tables 3 and 4 give, for each of the 9 stands, the values of mean and standard deviation of the five colour parameters L*, a*, b*, C* and H*. Table 3 presents the values obtained for the only 10 peripheral positions labelled PW (10 per log); as a comparison table 4 gives the mean of all radial positions in the bottom logs. Mean values of L* are higher while those of a* and b* are lower, when peripheral values are compared to all measured values. Additionaly standard deviations are all increased (more than 60% for L* and about 75% for a*) when comparing peripheral and all values. This strong variation is due to the occurrence of red heartwood that produce a darker and redder wood.

Figure 6 illustrates these variations of colour for L* only on the peripheral positions. Each vertical bar corresponds to the bottom log of one of the 87 trees tested, with the black mark giving the mean and the half-length of the bar the standard deviation. The trees are grouped by stands, separated by vertical dotted lines. The means and standard deviations of the 9 stands are grouped on the extreme right of the graph.

These graph and tables evidence a highly significant stand effect (at the 0.1% level). For instance the wood from stands D1, F1, G1 and S1 appears lighter, less red and less yellow than that from stands A1, A2, S2. There is also a clear tree effect within some stands, made apparent in the figure (e.g., stands A1, S1). Table 5 presents the ANOVA.

**Differences of the colour parameters among positions in the stem**

The colour parameters of boards coming from the north (N) and the south (S) directions of the same log were not significantly different even at the 5% level. We also compared logs situated in the height of 4m (bottom) and 9m (top) in the stem. When only the peripheral wood values were considered, the difference between colour parameters was not significant even at the 5% level; figure 7 shows the absence of log-by-log colour differences.

# Conclusion

Different variance analysis have put in evidence both a stand and a tree effect on colour variations for beech wood. At the first level this is true for the occurrence of red heartwood inside of the logs. Although the number of stands is small, it seems that stand effect is a very important parameter for red heartwood development. But stand effect in our case is a complex mixture of soil, climate, age and silviculture management. When putting aside the red heartwood there is still both a highly significant effect of stand and tree on peripheral wood colour. Besides, the differences remain very low inside one tree from bottom to top, north to south, and from outside to inside (except very near the pith or in case of red heartwood). Thus it is possible to sort or select on colour components rather easily at early stage from periphery, or from increment cores [5]. This can be used either by foresters or by industry depending on the objective.

Red heartwood is strongly darker, redder and more yellow than peripheral wood. Including or not the red heartwood there exists very strong relationships between colour components of beech wood, mainly for the couples $L^*/a^*$ and $a^*/b^*$.

In order to use also beech red heartwood it seems necessary to sort it away and to use it separately because the colour differences are very high and remain high after heating.

# Acknowledgements

This work was performed in the frame of the contract FAIR-98-3606 "Stresses in beech" supported by the European Commission [1], as well as with the financial support of CNRS-K.C. Wong post-doctoral program.

## Legends of figures

Figure 1:    Schematic localisation of colour measurement points for boards

Figure 2:    Beech colour compared to that of other hardwoods

Figure 3:    Examples of variation patterns from pith to bark of colour parameters: (a) type RR (a Swiss tree); (b) type NN (a German tree); (c) type RN (a Danish tree)

Figure 4:    Relationships between peripheral wood and heartwood for L*, a* and b*, in mean per boards (74 points for NRH and 97 points for RH)

Figure 5:    Relationship between L*/a*, L*/b* and a*/b*, separating peripheral wood (PW), red heartwood (RHW) and non red heartwood points (mean for 342 boards)

Figure 6:    Mean colour for L*, a* and b*, for peripheral wood values per tree

Figure 7:    Comparison between bottom and top logs (16 trees)

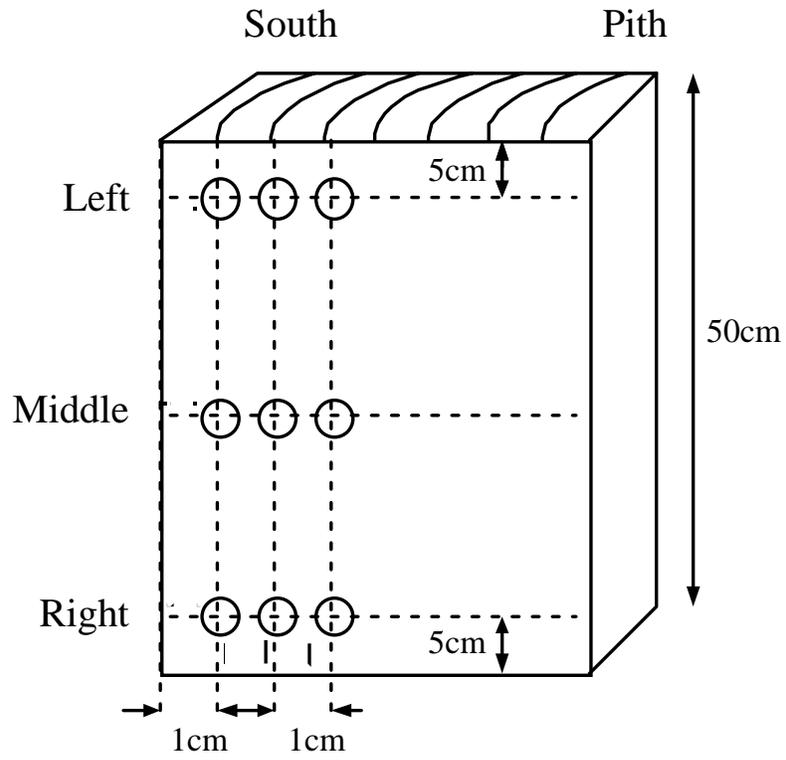

Figure 1: Schematic localisation of measurement points for boards

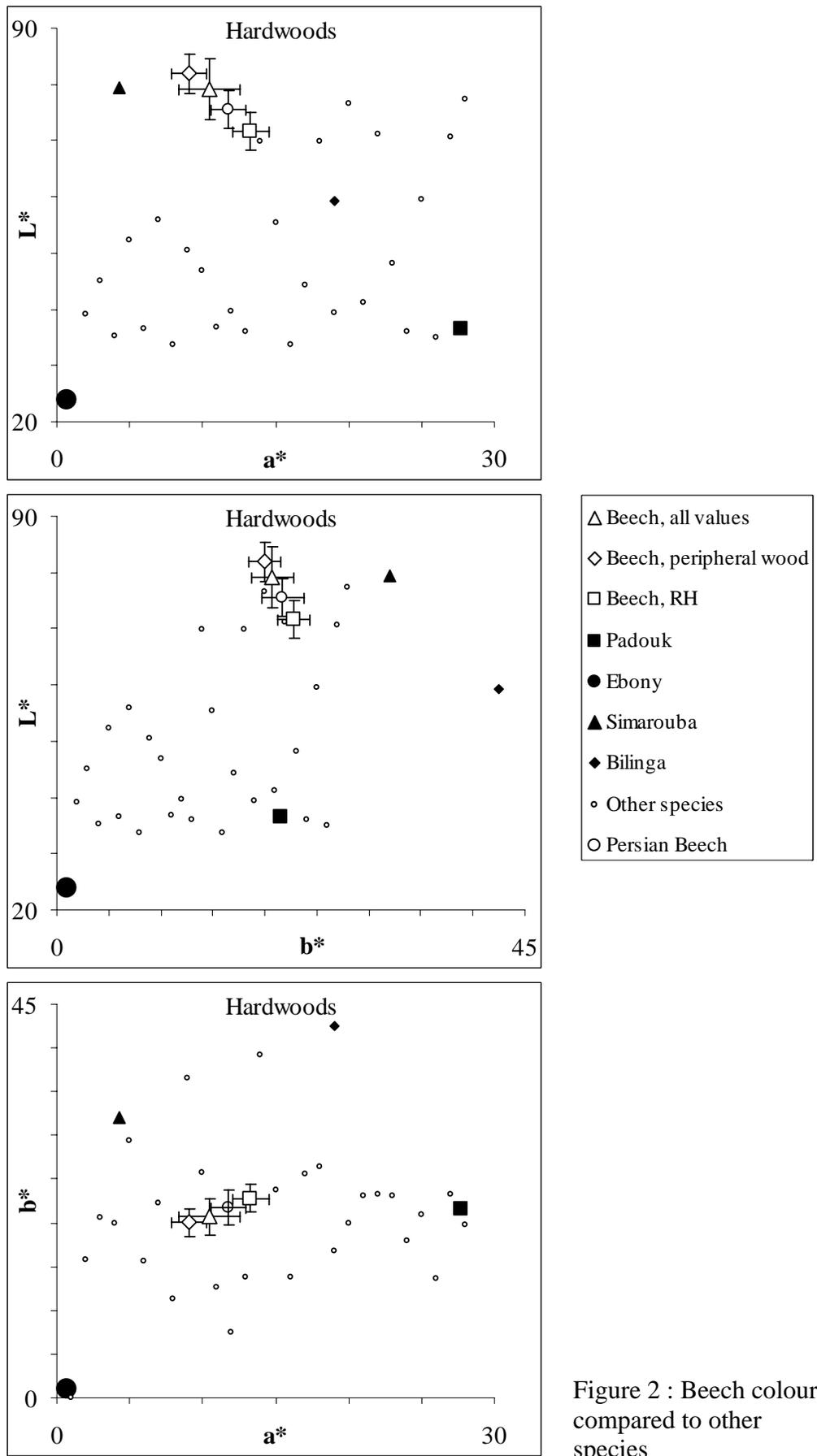

Figure 2 : Beech colour compared to other species

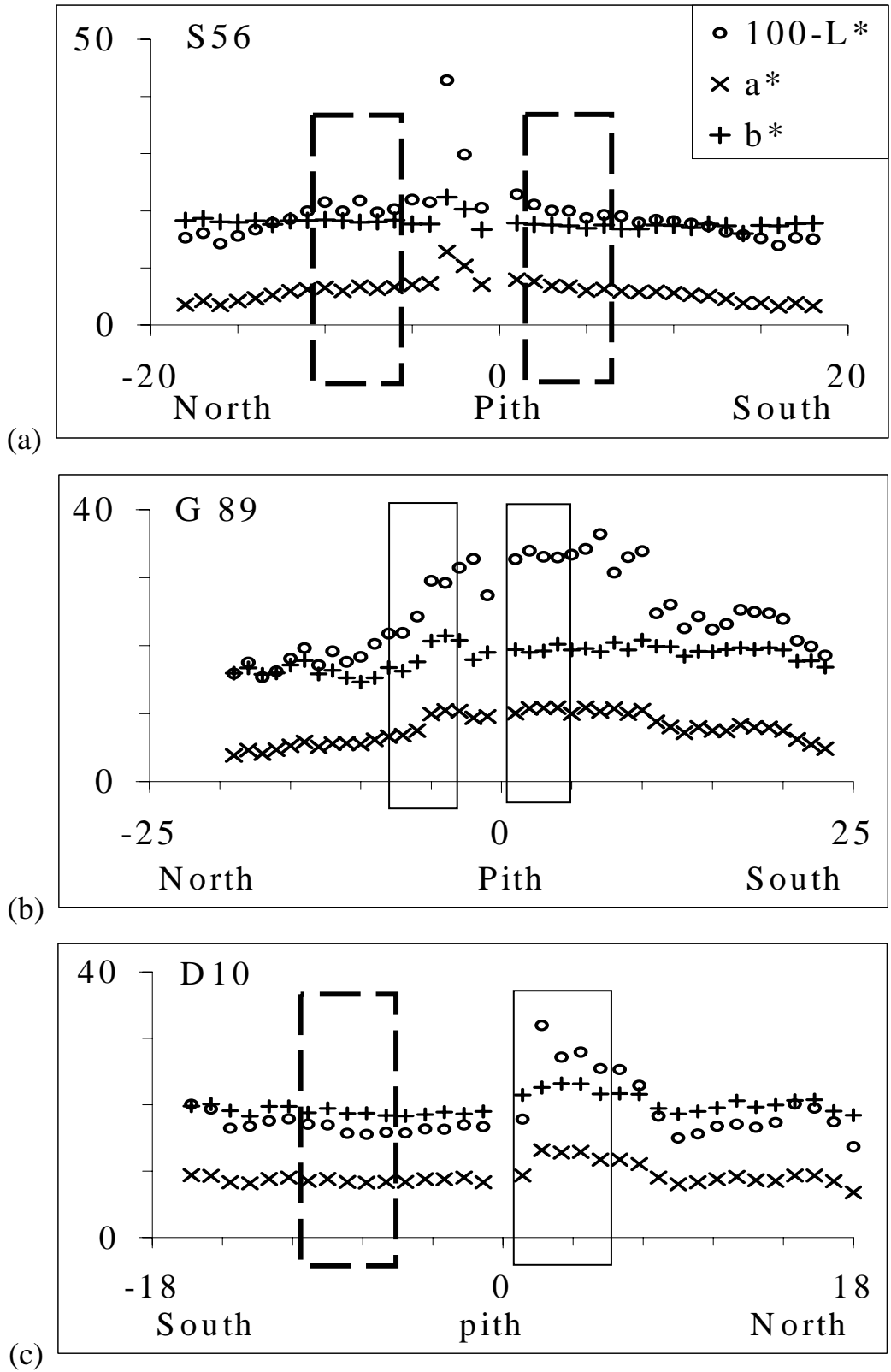

Figure 3: Examples of variation patterns from pith to bark of colour parameters:
(a) type I (a Swiss tree); (b) type II (a German tree); (c) type III (a Danish tree)

⌐ ⌐ Non Red Heartwood          ☐ Red Heartwood

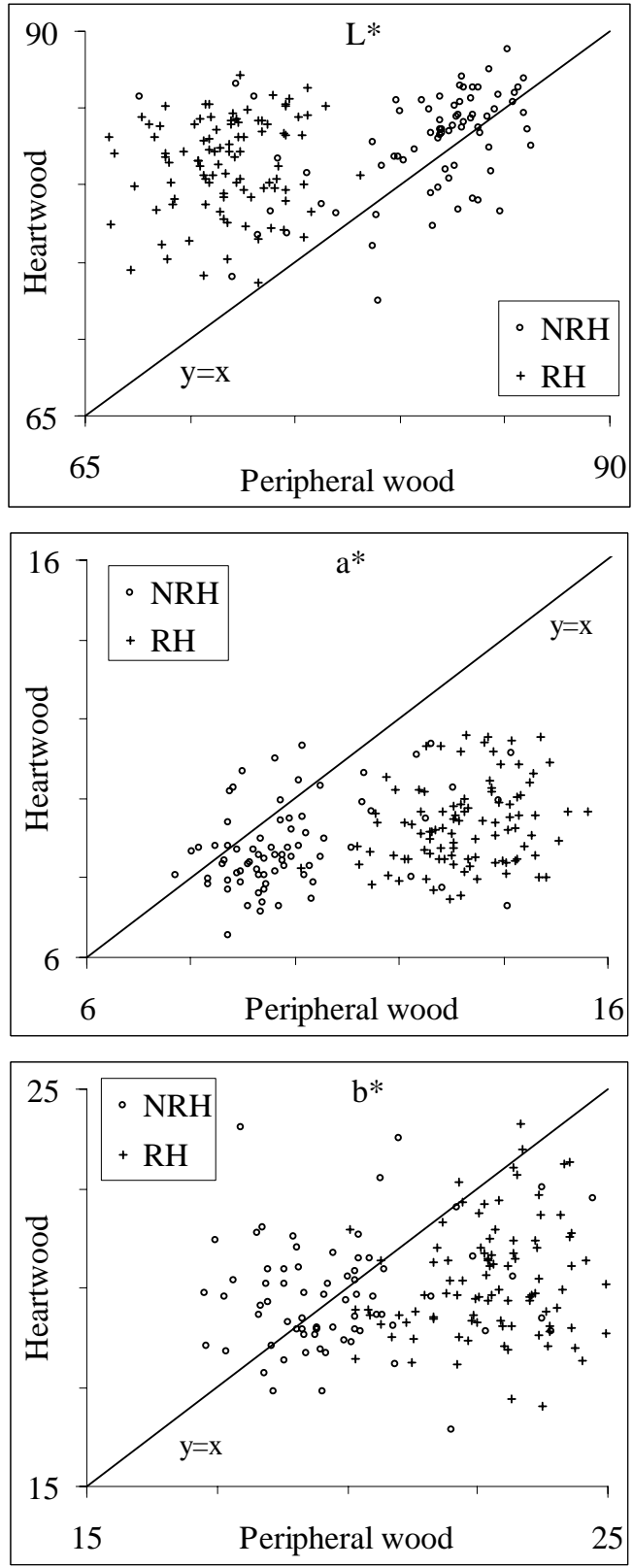

Figure 4: Relationship between peripheral wood and heartwood for L*, a* and b* in mean per boards (74 points for Non Red Heartwood and 97 points for Red Heartwood)

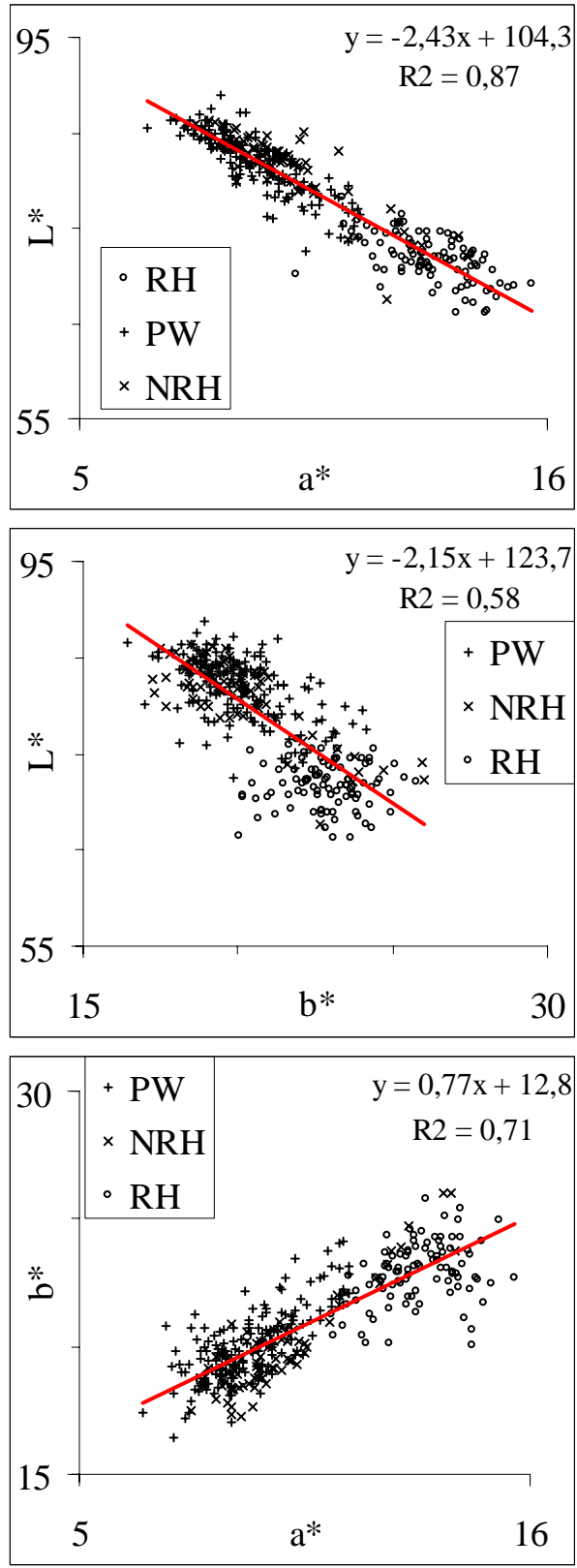

Figure 5: relationship between L*/a*; l*/b* and a*/b* in mean per board (342 boards)

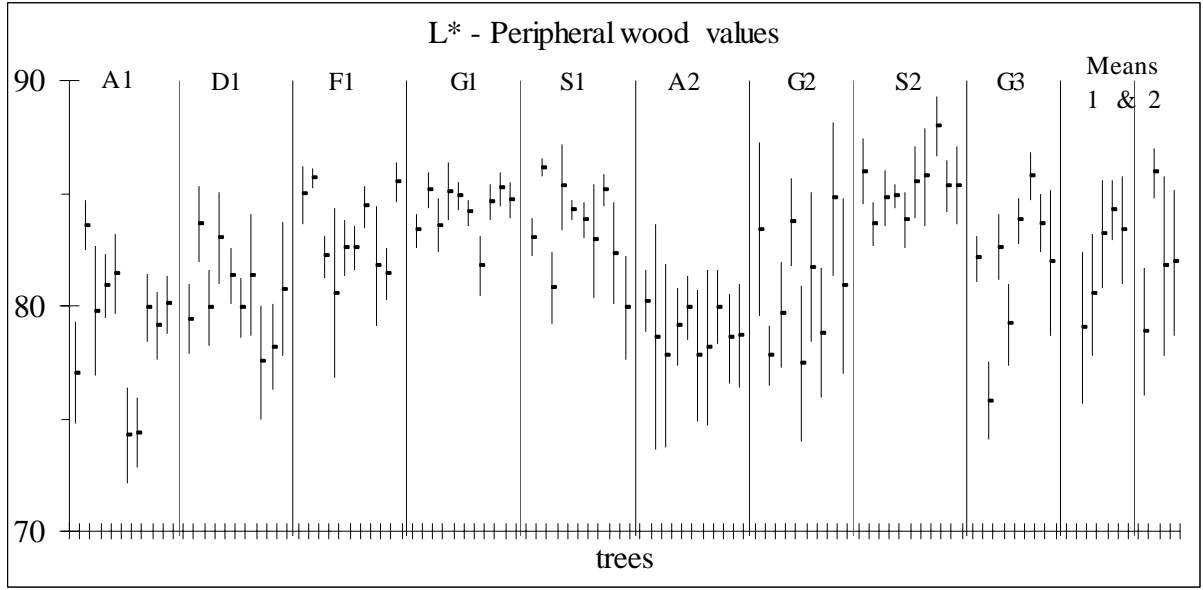

Figure 6: Mean per tree of L* for peripheral wood values

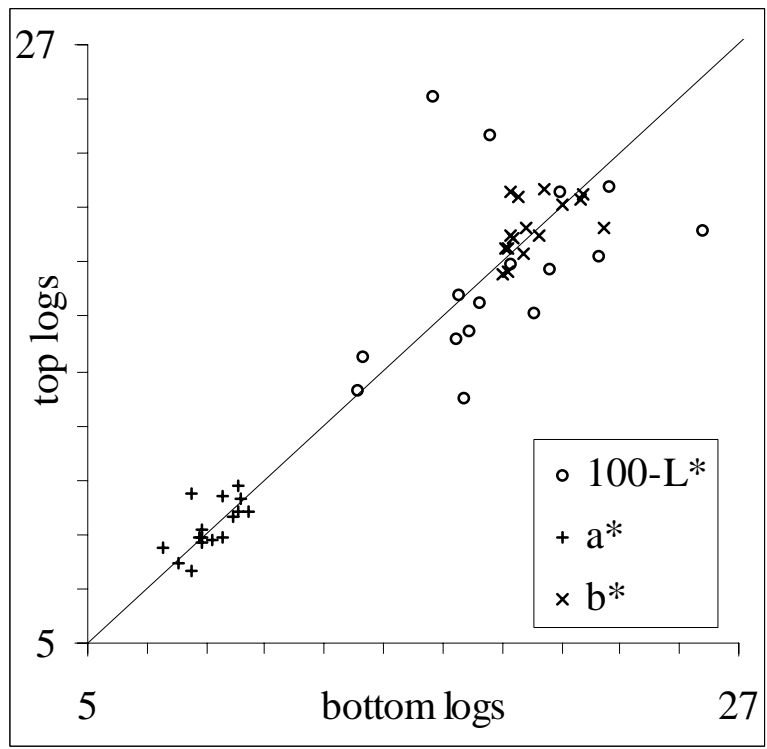

Figure 7: Comparison between bottom and top logs for colour parameters (16 trees)

Table 1: Tree main data

| Stand designation | tree nb | log age | H (m) | BH (m) | DBH (cm) | Board type | Board % RHW | Stand designation | tree nb | log age | H (m) | BH (m) | DBH (cm) | Board type | Board % RHW |
|---|---|---|---|---|---|---|---|---|---|---|---|---|---|---|---|
| A1 (Austria) high forest | 1 | 139 | 31.5 | 23 | 57 | R | 75.5 | A2 (Austria) mountain forest, with slope | 51 | 123 | 31 | 17 | 53 | R* | 26.7 |
| | 2 | 130 | 31.5 | 24 | 55 | R | 57.8 | | 52 | 126 | 31 | 16 | 51 | R | 45.3 |
| | 9 | - | 30.5 | 20 | 57 | R | 67.3 | | 82 | 131 | 35 | 22 | 50 | RN | 21.8 |
| | 10 | 131 | 33.5 | 23 | 52 | R | 50.0 | | 86 | 124 | 33 | 19 | 47 | R | 47.6 |
| | 16 | 123 | 35 | 23.5 | 57 | R | 41.3 | | 89 | 121 | 34 | 22 | 47 | R | 37.2 |
| | 18 | 112 | 30 | 23 | 51 | N | 15.9 | | 90 | 123 | 33 | 17 | 53 | RN | 17.4 |
| | 23 | - | 33 | 21 | 54 | R | 35.7 | | 94 | 105 | 33 | 21 | 49 | N | 12.8 |
| | 27 | 150 | 32 | 19 | 56 | R | 36.4 | | 95 | - | 32 | 17 | 50 | R | 30.8 |
| | 29 | 151 | 34 | 20 | 57 | R | 62.2 | | 96 | 118 | 31 | 14 | 49 | R | 33.3 |
| | 30 | - | 35 | 18 | 63 | R | 36.5 | | 99 | 121 | 31 | 18 | 48 | R | 39.0 |
| Mean A1 | - | 134 | 32.6 | 21.5 | 56 | - | 47.9 | Mean A2 | - | 121 | 32 | 18.3 | 49.7 | - | 31.2 |
| S1 (Switzerland) high forest | 1 | 126 | 30.8 | 16.9 | 63 | R | 49.0 | S2 (Switzerland) mountain forest, with slope | 56 | 159 | 39 | 22 | 49 | N | 0.0 |
| | 11 | 55 | 36.9 | 9.8 | 55 | N | 7.1 | | 57 | - | 41 | 25 | 54 | N | 17.8 |
| | 14 | 75 | 23.7 | 12.7 | 55 | RN | 27.7 | | 63 | 158 | 37 | 25 | 60 | N | 0.0 |
| | 20 | 73 | 30.9 | 19.2 | 59 | N | 6.4 | | 64 | 158 | 37 | 19 | 68 | R | 29.5 |
| | 25 | - | 36.3 | 22.2 | 61 | RN | 27.7 | | 65 | 156 | 37 | 26 | 54 | N* | 28.6 |
| | 26 | 100 | 36.3 | 18.2 | 58 | N | 0.0 | | 74 | 158 | 38 | 14 | 63 | RN | 23.2 |
| | 28 | 97 | 33.7 | 20.6 | 51 | N | 0.0 | | 78 | 153 | 42 | 26 | 56 | RN | 22.0 |
| | 33 | 115 | 35.9 | 18.2 | 64 | N | 0.0 | | 80 | 153 | 42 | 24 | 63 | RN | 18.5 |
| | 37 | 118 | 38 | 22 | 52 | N | 23.3 | | 85 | 158 | 40 | 22 | 59 | N | 14.7 |
| | 39 | - | 37.4 | 20.8 | 52 | RN | 27.9 | | 86 | - | 39 | 16 | 59 | RN | 21.2 |
| Mean S1 | - | 95 | 34 | 18.1 | 56.9 | - | 16.9 | Mean S2 | - | 157 | 39 | 21.9 | 58.5 | - | 17.6 |
| G1 (Germany) high forest, narrow spacing | 3 | 125 | 34.3 | 19.8 | 53 | R | 30.6 | G2 (Germany) high forest, large spacing | 57 | 117 | 38 | 32 | 60 | RN | 25.2 |
| | 4 | 123 | 40.8 | 23.5 | 58 | N | 14.6 | | 59 | 111 | 36 | 19 | 62 | N | 0.0 |
| | 6 | 123 | 35.9 | 22.1 | 60 | R | 50.0 | | 64 | 116 | 36 | 22 | 61 | RN | 23.0 |
| | 7 | 125 | 40 | 19.6 | 54 | R | 53.2 | | 70 | 121 | 37 | 24 | 69 | R | 39.6 |
| | 9 | 121 | 39.1 | 14.9 | 60 | N | 0.0 | | 73 | 120 | 42 | 24 | 71 | R* | 12.7 |
| | 11 | 120 | 33.8 | 16.2 | 55 | R | 35.3 | | 79 | 119 | 42 | 22 | 83 | **Not cut** | |
| | 13 | 111 | 31.1 | 14 | 51 | N | 23.8 | | 89 | - | 37 | 20 | 55 | R | 40.4 |
| | 15 | 116 | 35 | 19.9 | 51 | RN | 27.5 | | 94 | 111 | 43 | 21 | 69 | N | 0.0 |
| | 19 | 121 | 36.1 | 21.4 | 53 | R | 35.4 | | 97 | - | 38 | 23 | 55 | R | 53.0 |
| | 20 | 117 | 33.6 | 19 | 57 | R | 34.0 | Mean G2 | - | 116 | 39 | 23 | 65 | - | 24.2 |
| Mean G1 | - | 120 | 36 | 19 | 55.5 | - | 30.4 | G3 (Germany) middle forest large spacing | 107 | 174 | 39 | 25 | 73 | R | 39.2 |
| D1 (Denmark) high forest | 3 | 101 | 42 | 27 | 64 | R | 28.6 | | 113 | 166 | 34 | 21 | 73 | R | 56.8 |
| | 7 | 98 | 33 | 19.2 | 57 | N | 0.0 | | 120 | 186 | 37 | 24 | 85 | R | 49.3 |
| | 8 | 101 | 31.5 | 18 | 63 | N | 9.4 | | 122 | 177 | 41 | 25 | 75 | R | 30.2 |
| | 10 | 88 | 36 | 18.6 | 48 | RN | 11.1 | | 125 | 171 | 37 | 23 | 84 | RN | 22.1 |
| | 11 | 93 | 37.2 | 19.2 | 66 | R | 35.8 | | 131 | 141 | 30 | 18 | 64 | R | 38.5 |
| | 12 | 106 | 37.5 | 22.5 | 58 | R | 25.0 | | 133 | - | 33 | 19 | 67 | RN | 17.6 |
| | 32 | 107 | 32.1 | 18 | 61 | N | 17.6 | | 150 | 142 | 31 | 14 | 77 | R | 26.7 |
| | 35 | 102 | 39.3 | 23.4 | 67 | R | 34.6 | Mean G3 | - | 165 | 35 | 21.1 | 74.8 | - | 35.1 |
| | 42 | 106 | 33 | 19.5 | 63 | R | 52 | Mean All | - | 122 | 36 | 20.6 | 58.8 | - | 26 |
| | 45 | 99 | 33.9 | 19.5 | 59 | N | 23.4 | | | | | | | | |
| Mean D1 | - | 99 | 35.8 | 20.6 | 60.5 | - | 20.6 | | | | | | | | |
| F1 (France) high forest | 2 | - | 35 | 20 | 52 | N | 17.6 | | | | | | | | |
| | 5 | 96 | 34 | 22 | 51 | N | 0.0 | | | | | | | | |
| | 6 | 96 | 37 | 24 | 56 | N | 17.4 | | | | | | | | |
| | 12 | 104 | 35 | 20 | 59 | N | 0.0 | | | | | | | | |
| | 20 | 95 | 38 | 25 | 54 | N | 0.0 | | | | | | | | |
| | 21 | - | 40 | 20 | 57 | N | 10.0 | | | | | | | | |
| | 22 | 105 | 36 | 26 | 62 | N | 0.0 | | | | | | | | |
| | 28 | 100 | 35 | 24 | 64 | N | 0.0 | | | | | | | | |
| | 33 | - | 39 | 24 | 57 | R | 33.3 | | | | | | | | |
| | 43 | 97 | 32 | 20 | 54 | R | 34.0 | | | | | | | | |
| Mean F1 | - | 99 | 36.1 | 22.5 | 56.6 | - | 11.2 | | | | | | | | |

Log age: number of year rings measured on the bottom log used for the study; the total age of the tree is about 20 years more than the bottom age; DBH: Diameter at breast height, H: tree height; BH: "base" height or distance from soil to living crown; Board Type: Board considered as having Red Heartwood (R) or not (N) and (RN) have the two kinds of boards (three boards has been excluded); "board % RHW" percentage of red heartwood (RH width / board width) "-": missing age data and "*": one of the board is excluded. Log 79 was too large to be sawn.

Table 2: data for the 34 top logs (see legend in table 1)

| Stand designation | tree nb | log age | Board Type | % RHW | Stand designation | tree nb | log age | Board Type | % RHW |
|---|---|---|---|---|---|---|---|---|---|
| A1 | 16 | 107 | R | 35.4 | A2 | 86 | 102 | R | 46.3 |
|    | 18 | 100 | N | 0.0  |    | 90 | -   | RN | 20.7 |
| S1 | 37 | 97  | N | 0.0  | S2 | 56 | 145 | N  | 0.0  |
|    | 39 | 117 | R | 27.6 |    | 80 | 141 | R  | 27.8 |
| G1 | 3  | 114 | R | 30.4 | G2 | 73 | 109 | R  | 27.5 |
|    | 13 | 99  | N | 0.0  |    | 79 | 119 | R  | 43.1 |
| D1 | 3  | 83  | R | 35.7 | G3 | 107| 166 | R  | 36.1 |
|    | 45 | 82  | N | 0.0  |    | 113| 168 | R  | 56.6 |
| F1 | 21 | -   | N | 9.0  |    |    |     |    |      |

Table 3: mean and standard deviation per stand for peripheral wood values (10 points per tree)

|    |      | A1    | D1    | F1    | G1    | S1    | A2    | G2    | S2    | G3    | All   |
|----|------|-------|-------|-------|-------|-------|-------|-------|-------|-------|-------|
| L* | Mean | 79,03 | 80,49 | 83,15 | 84,24 | 83,36 | 78,87 | 80,89 | 85,34 | 81,92 | 81,95 |
|    | Sd   | 3,33  | 2,68  | 2,39  | 1,36  | 2,42  | 2,85  | 3,89  | 1,73  | 3,23  | 3,48  |
| a* | Mean | 9,60  | 9,44  | 8,76  | 8,30  | 8,55  | 10,36 | 9,61  | 8,00  | 9,42  | 9,10  |
|    | Sd   | 0,93  | 1,02  | 0,71  | 0,60  | 1,02  | 0,88  | 1,27  | 0,69  | 1,29  | 1,19  |
| b* | Mean | 19,82 | 20,15 | 19,52 | 18,91 | 19,84 | 21,06 | 20,33 | 19,59 | 21,27 | 20,02 |
|    | Sd   | 1,32  | 1,46  | 1,05  | 0,92  | 1,51  | 1,62  | 2,05  | 0,86  | 1,46  | 1,55  |
| C  | Mean | 22,03 | 22,26 | 21,39 | 20,66 | 21,61 | 23,48 | 22,5  | 21,16 | 23,28 | 22    |
|    | Sd   | 1,54  | 1,73  | 1,19  | 1,04  | 1,72  | 1,78  | 2,31  | 0,93  | 1,73  | 1,81  |
| H  | Mean | 64,2  | 64,95 | 65,85 | 66,32 | 66,73 | 63,79 | 64,73 | 67,79 | 66,17 | 65,62 |
|    | Sd   | 1,23  | 1,14  | 1,16  | 0,96  | 1,59  | 1,30  | 1,76  | 1,62  | 2,21  | 1,91  |
| Nb values |  | 100 | 100 | 100 | 100 | 100 | 100 | 80 | 100 | 80 | 860 |

Table 4: mean and standard deviation per stand for all radii values measured on bottom logs

|    |      | A1    | D1    | F1    | G1    | S1    | A2    | G2    | S2    | G3    | All   |
|----|------|-------|-------|-------|-------|-------|-------|-------|-------|-------|-------|
| L* | Mean | 75,70 | 79,50 | 81,20 | 80,38 | 81,68 | 76,43 | 77,21 | 81,95 | 77,00 | 79,05 |
|    | Sd   | 5,32  | 4,35  | 4,32  | 5,41  | 4,66  | 4,34  | 5,30  | 4,67  | 5,87  | 5,45  |
| a* | Mean | 11,10 | 10,15 | 9,59  | 9,72  | 9,23  | 11,68 | 11,32 | 9,83  | 11,49 | 10,44 |
|    | Sd   | 2,00  | 1,75  | 1,50  | 1,89  | 1,56  | 1,88  | 2,07  | 2,01  | 2,25  | 2,09  |
| b* | Mean | 21,03 | 20,74 | 19,91 | 19,86 | 19,80 | 21,99 | 21,72 | 20,13 | 21,50 | 20,72 |
|    | Sd   | 2,03  | 2,05  | 1,86  | 1,88  | 1,65  | 1,90  | 2,37  | 1,33  | 1,92  | 2,06  |
| C  | Mean | 23,80 | 23,11 | 22,11 | 22,14 | 21,87 | 24,92 | 24,52 | 22,45 | 24,43 | 23,23 |
|    | Sd   | 2,66  | 2,58  | 2,25  | 2,46  | 2,06  | 2,45  | 2,93  | 1,95  | 2,52  | 2,67  |
| H  | Mean | 62,38 | 64,09 | 64,38 | 64,13 | 65,12 | 62,18 | 62,65 | 64,20 | 62,09 | 63,48 |
|    | Sd   | 2,51  | 1,87  | 2,04  | 2,59  | 2,43  | 2,43  | 2,64  | 3,55  | 3,66  | 2,92  |
| Nb values |  | 429 | 439 | 423 | 444 | 416 | 400 | 369 | 458 | 461 | 3839 |

Table 5 Analysis of variance for the 3 colour parameters L*, a*, b*

|    | Source of variation | df | SS | MS | F | P-value |
|----|---|---|---|---|---|---|
| L* | Among stands  | 8  | 412   | 51.5  | 11.83 | 8E-11 |
|    | Within stands | 77 | 335.2 | 4.354 |       |       |
|    | Total         | 85 | 747.3 |       |       |       |
| a* | Among stands  | 8  | 45.37 | 5.671 | 11.39 | 2E-10 |
|    | Within stands | 77 | 38.32 | 0.498 |       |       |
|    | Total         | 85 | 83.69 |       |       |       |
| b* | Among stands  | 8  | 41.68 | 5.21  | 4.423 | 2E-04 |
|    | Within stands | 77 | 90.71 | 1.178 |       |       |
|    | Total         | 85 | 132.4 |       |       |       |
| Critical value for F (0.1% level) | | | | 3.723 | | |